\documentclass[
  ,final            
  ]
  {aipproc}

\layoutstyle{8x11double}

\begin{document}

\title{The CLEO-c Research Program}

\author{David Asner - for the CLEO collaboration}{
  address={University of Pittsburgh, Department of Physics and Astronomy, 3951 O'Hara St, Pittsburgh PA, 15260, USA}
}

\begin{abstract}

The CLEO-c research program will include studies of leptonic,
semileptonic
and hadronic charm decays, searches for exotic and gluonic matter, and test for
physics beyond the Standard Model. 
In the summer of 2003 the experiment and the CESR 
accelerator were modified to operate at center-of-mass energies between 3 and 5~GeV.
Data at the $\psi(3770)$ resonance were recorded with 
the CLEO-c detector in September 2003.
beginning a new era in the exploration of the charm sector.

\end{abstract}

\maketitle


\section{Introduction}

The CLEO-c physics program~\cite{bib:cleoc} includes a variety of measurements
that will improve the understanding of Standard Model processes
as well as provide the opportunity to probe physics that lies beyond the
Standard Model. 
The primary components of this program are measurement of absolute
branching ratios for charm mesons with a precision of the order of $1-2\%$,
determination of charm meson decay constants and of
the CKM matrix elements $\mid V_{cs} \mid$ 
and $\mid V_{cd} \mid$ at the $1 - 2\%$ level and
investigation of processes in charm decays that are
highly suppressed within the Standard Model. 
A 10 nb$^{-1}$ cross section for $e^+e^- \to D{\overline D}$ is assumed throughout ref.~\cite{bib:cleoc}.


Beginning in 2003 the CESR accelerator will be operated at
center-of-mass energies corresponding to  $\sqrt{s}
\sim 3770$~MeV ($\psi^{\prime\prime}$), $\sqrt{s} \sim 4140$~MeV and $\sqrt{s} \sim 3100$~MeV ($J/\psi$).
The luminosity over this energy range will range from $5 \times 10^{32}
cm^{-2} s^{-1}$ down to about $1 \times 10^{32} cm^{-2} s^{-1}$ 
yielding 3~fb$^{-1}$ each at the $\psi^{\prime\prime}$ and at
$\sqrt{s} \sim 4140$~MeV above $D_s \bar{D_s}$ threshold and 1~fb$^{-1}$ at the
$J/\psi$. These integrated luminosities correspond to samples of 1.5 million
$D_s \bar{D_s}$ pairs, 30 million $D \bar{D}$ pairs and one billion $J/\psi$
decays~\cite{bib:cleoc}. These datasets will exceed those of the
Mark III experiment by factors of 480, 310 and 170, respectively.
Table \ref{tab:runplan} summarizes the run plan.

\par
From fall 2001 to spring 2003 CLEO collected a total of 4~fb$^{-1}$ of
data on the $\Upsilon(1S)$, $\Upsilon(2S)$, $\Upsilon(3S)$ and $\Upsilon(5S)$
which is currently under analysis.
These data sets will
increase the available $b\bar{b}$ bound state data by more than an order of
magnitude.

\begin{table}
\begin{tabular}{lll}
\hline
\multicolumn{1}{l}{\bf Resonance}
  & \multicolumn{1}{l}{\bf Anticipated}
  & \multicolumn{1}{l}{\bf Reconstructed}\\
&  \multicolumn{1}{l}{\bf Luminosity} &  \multicolumn{1}{l}{\bf Events}   \\
\hline
 $\psi(3770)$             & $\sim$ 3~fb$^{-1}$ & 30M $D\bar{D}$\\
 $\sqrt{s} \sim 4140$~MeV & $\sim$ 3~fb$^{-1}$ & 1.5M $D_s\bar{D_s}$\\
 $\psi(3100)$             & $\sim$ 1~fb$^{-1}$ & 60M radiative $J/\psi$  \\
\hline
\end{tabular}
\caption{The 3-year CLEO-c run plan~\cite{bib:cleoc}}
\label{tab:runplan}
\end{table}

Only modest hardware modifications are required for low energy operation.
The transverse cooling of the CESR beams will be enhanced by 16 meters of superconducting wiggler magnets. Half of the full complement of 12 wigglers were installed 
in summer 2003 with the additional 6 wigglers scheduled for installation in 2004. 
The CLEO III silicon vertex detector was replaced by a
small, low mass inner drift chamber.
The solenoidal field was
reduced from 1.5~T to 1.0~T. No other modifications are planned.


\section{Physics Program}

The following sections will outline the CLEO-c physics program.
The first section will focus on the Upsilon spectroscopy, the second section
will describe the charm decay program, the third section
will give an overview about the exotic and gluonic matter studies and the last
section will descibe the oportunities to probe 
physics beyond the Standard Model.

\subsection{Upsilon Spectroscopy \label{sec:upsilon}}

The only established $b\bar{b}$ states below $B\bar{B}$ threshold are the three
vector triplet $\Upsilon$ resonances ($^3S_1$) and the six $\chi_b$ and
$\chi'_b$ (two triplets of $^3P_J$) that are accessible from these parent
vectors via E1 radiative transitions.
CLEO will address a
variety of outstanding physics issues with the data samples
at the $\Upsilon(1S)$, $\Upsilon(2S)$ and $\Upsilon(3S)$, 



\noindent {\bf Searches for the $\eta_b$ and $h_b$:}
The $\eta_b$ is the ground state of $b\bar{b}$. Most present theories 
~\cite{bib:EFI01-10} indicate
the best approach would be the hindered M1 transition from the $\Upsilon(3S)$,
with which CLEO might have a signal of $5 \sigma$ significance in 1~fb$^{-1}$
of data. In the case of the $h_b$, CLEO established an upper limit of
${\cal B}(\Upsilon(3S) \rightarrow \pi^+ \pi^- h_b)$ $<$ 0.18\% at 90\%
confidence level~\cite{bib:hb}. This result, based on $\sim 110$~pb$^{-1}$, 
already
tests some theoretical predictions~\cite{bib:hb2,bib:hb3,bib:hb4} for this transition which range
from $< 0.01 - 1.0\%$. 

\noindent {\bf Observation of $1^3D_J$ states:}
The $b\bar{b}$ system is unique as it has states with L = 2 that lie below
the open-flavor threshold. These states have been of considerable theoretical
interest, as indicated by many predictions of the center-of-gravity of the
triplet and by a recent review~\cite{bib:EFI01-14}. 
In an analysis of the $\Upsilon(3S)$ CLEO data sample the 
$\Upsilon(1^3D_2)$ state could already be observed in the
four-photon cascade
$\Upsilon(3S) \rightarrow \gamma_1\chi'_b
\rightarrow \gamma_1\gamma_2 \Upsilon(^3D_J) \rightarrow \gamma_1\gamma_2\gamma_3
\chi_b \rightarrow \gamma_1\gamma_2\gamma_3\gamma_4\ell^+\ell^-$.
The mass of the $\Upsilon(1^3D_2)$ state is determined to be
$10161.1 \pm 0.6 \pm 1.6$~MeV/$c^2$~\cite{bib:lp2003}.

\noindent {\bf Glueball candidates in radiative $\Upsilon(1S)$ decays:}
Signals for glueball candidates
in radiative $J/\psi$ decay - a glue-rich environment - might be observed in
radiative $\Upsilon(1S)$ decays.
Naively
one would expect the exclusive radiative decay to be suppressed in $\Upsilon$
decay by a factor of roughly 40, which implies product branching fractions for
$\Upsilon$ radiative decay of $\sim 10^{-6}$. With 1~fb$^{-1}$ of data and
efficiencies of around 30\% one can expect $\sim$ 10 events in each of the
exclusive channels, which would be an important confirmation of the $J/\psi$
studies.

\subsection{Charm Decays \label{sec:charm}}
The observable properties of the charm mesons are determined by the strong
and weak interactions. As a result, charm mesons can be used as a laboratory
for the studies of these two fundamental forces. Threshold charm experiments
permit a series of measurements that enable direct study of the weak
interactions of the charm quark, as well as tests of our theoretical technology
for handling the strong interactions.

\noindent{\bf Leptonic Charm Decays:}
Measurements of leptonic decays in CLEO-c will benefit from the fully 
tagged $D^+$ and $D_s$ decays available at the $\psi(3770)$ and at 
$\sqrt{s} \sim 4140$~MeV. The leptonic decays $D_s \rightarrow \mu\nu$
are detected in tagged events by observing a single charged track of the
correct sign, missing energy, and a complete accounting of the residual
energy in the calorimeter. The clear definition of the initial state, the
cleanliness of the tag reconstruction, and the absence of additional 
fragmentation tracks make this measurement straightforward and essentially
background-free. This will enable measurements of the poorly known
leptonic decay rates for $D^+$ and $D^+_s$ to a precision of 3 - 4\% and will allow
the validation of theoretical calculations of the decay constants
$f_{D}$ and $f_{D_s}$ at the 1 - 2 \% level. 
Table \ref{tab:decayconsts} summarizes
the expected precision in the decay constant measurements.

\begin{table}[ht]
\begin{tabular}{lcc}
\hline
    \multicolumn{1}{l}{\bf }
  & \multicolumn{2}{c}{\bf Decay Constant Error \%} \\
    \multicolumn{1}{l}{\bf Decay Mode}
  & \multicolumn{1}{l}{\bf PDG 2000}
  & \multicolumn{1}{l}{\bf CLEO-c~\cite{bib:cleoc}}   \\
\hline
$D^+ \rightarrow \mu^+ \nu$ ($f_D$)     & Upper Limit  & 2.3  \\
$D^+_s \rightarrow \mu^+ \nu$   ($f_{D_s}$) & 17           & 1.7  \\
$D^+_s \rightarrow \tau^+ \nu$  ($f_{D_s}$) & 33           & 1.6  \\
\hline
\end{tabular}
\caption{Expected decay constants errors for leptonic decay modes}
\label{tab:decayconsts}
\end{table}

\noindent{\bf Semileptonic Charm Decays:}
The CLEO-c program will provide a large set of precision measurements in the
charm sector against which the theoretical tools needed to extract CKM matrix
information precisely from heavy quark decay measurements will be tested and
calibrated.
\par
CLEO-c will measure the branching ratios of many exclusive semileptonic modes,
including
$D^0 \rightarrow K^- e^+ \nu$,
$D^0 \rightarrow \pi^- e^+ \nu$,
$D^0 \rightarrow K^{-} e^+ \nu$,
$D^+ \rightarrow \bar{K}^{0} e^+ \nu$,
$D^+ \rightarrow \pi^0 e^+ \nu$,
$D^+ \rightarrow \bar{K}^{0*} e^+ \nu$,
$D^+_s \rightarrow \phi e^+ \nu$ and
$D^+_s \rightarrow \bar{K}^{0*} e^+ \nu$.
The measurement in each case is based on the use of tagged events where the
cleanliness of the environment provides nearly background-free signal samples,
and will lead to the determination of the CKM matrix elements
$\mid V_{cs} \mid$ and $\mid V_{cd} \mid$ with a precision level of
1.6\% and 1.7\%, respectively. Measurements of the vector and axial vector form
factors $V(q^2)$, $A_1(q^2)$ and $A_2(q^2)$ will also be possible at the
$\sim$ 5\% level. Table \ref{tab:semileptonic} summarizes the expected fractional error
on the branching ratios.

\begin{table}[ht]
\begin{tabular}{lcc}
\hline
    \multicolumn{1}{l}{\bf }
  & \multicolumn{2}{c}{\bf BR fractional error \%} \\
    \multicolumn{1}{l}{\bf Decay Mode}
  & \multicolumn{1}{l}{\bf PDG 2000}
  & \multicolumn{1}{l}{\bf CLEO-c~\cite{bib:cleoc}}   \\
\hline
$D^0 \rightarrow K \ell \nu$    & 5        & 0.4                       \\
$D^0 \rightarrow \pi \ell \nu$  & 16       & 1.0                      \\
$D^+ \rightarrow \pi \ell \nu$  & 48       & 2.0                      \\
$D_s \rightarrow \phi \ell \nu$ & 25       & 3.1                       \\
\hline
\end{tabular}
\caption{Expected branching fractional errors for selected semileptonic decay modes}
\label{tab:semileptonic}
\end{table}

HQET provides a successful description of the lifetimes of charm hadrons and
of the absolute semileptonic branching ratios of the $D^0$ and $D_s$
~\cite{bib:bcp3}. Isospin invariance in the strong forces implies
$\Gamma_{SL}(D^0) \simeq \Gamma_{SL}(D^+)$ up to corrections of 
${\cal O}(\tan^2 \theta_C) \simeq 0.05$. Likewise, $SU(3)_{Fl}$ symmetry
relates $\Gamma_{SL}(D^0)$ and $\Gamma_{SL}(D_s^+)$, but a priori would
allow them to differ by as much as 30\%. However, HQET suggests that they
should agree to within a few percent. The charm threshold region 
is the best place to
measure absolute inclusive semileptonic charm branching ratios, in particular
${\cal B}(D_s \rightarrow X \ell\nu)$ and thus $\Gamma_{SL}(D_s)$.

\noindent{\bf Implications for CKM Triangle:}
%
The CLEO-c program of leptonic and semileptonic
measurements has two components: one of calibrating and validating
theoretical methods for calculating hadronic matrix elements, which can then be
applied to all problems in CKM extraction in heavy quark physics; and one of
extracting CKM elements directly from the CLEO-c data. The direct results
of CLEO-c are the precise determination of $\mid V_{cd} \mid$, 
$\mid V_{cs} \mid$, $f_D$,
$f_{D_s}$, and the semileptonic form factors. The precision knowledge
of the decay constants $f_D$ and $f_{D_s}$, together with the rigorous
calibration of theoretical techniques for calculating heavy-to-light
semileptonic form factors, are required for the direct extraction of CKM 
elements from CLEO-c. This also drives
the indirect results, namely the precision 
extraction
of CKM elements from experimental measurements of the $B_d$ mixing 
frequency, the $B_s$ mixing frequency, and the $B \rightarrow \pi\ell\nu$
decay rate measurements which will be performed by
BaBar, Belle, CDF, D0, BTeV, LHCb, ATLAS and CMS.
In Table \ref{tab:newCKM} the combined projections are presented~\cite{bib:cleoc}. In the
determination of the CKM elements $\mid V_{cd} \mid$ 
and $\mid V_{cs} \mid$ from $B$ and $B_s$
mixing $\mid V_{tb} \mid = 1$ is used. The tabulation also includes improvement
in the direct measurement of $\mid V_{tb} \mid$ 
expected from the Tevatron experiments~\cite{bib:tevatron1}. 

\begin{table}[ht]
\begin{tabular}{lll}
\hline
    \multicolumn{3}{c}{\bf Present Knowledge} \\
\hline
$\delta V_{ud}/V_{ud}=$ 0.1\% & $\delta V_{us}/V_{us}=$ 1\%   & 
$\delta V_{ub}/V_{ub}=$ 25\%  \\ 
$\delta V_{cd}/V_{cd}=$ 7\%   & $\delta V_{cs}/V_{cs}=$ 16\%  & 
$\delta V_{cb}/V_{cb}=$ 5\%   \\
$\delta V_{td}/V_{td} = 36\%$  & $\delta V_{ts}/V_{ts} = 39\%$  & 
$\delta V_{tb}/V_{tb} = 29\%$  \\ 
\hline
   \multicolumn{3}{c}{\bf After CLEO-c}   \\
\hline
$\delta V_{ud}/V_{ud}=$ 0.1\% &
$\delta V_{us}/V_{us}=$ 1\%   & $\delta V_{ub}/V_{ub}=$ 5\%   \\
$\delta V_{cd}/V_{cd}=$ 1\%   &
$\delta V_{cs}/V_{cs}=$ 1\%   & $\delta V_{cb}/V_{cb}=$ 3\%   \\
$\delta V_{td}/V_{td}=$ 5\%   &
$\delta V_{ts}/V_{ts}=$ 5\%   & $\delta V_{tb}/V_{tb}=$ 15\%  \\
\hline
\end{tabular}
\caption{CKM elements at present and after CLEO-c~\cite{bib:cleoc}}
\label{tab:newCKM}
\end{table}

\noindent{\bf Hadronic Charm Decays:}
The CLEO and ALEPH
experiments by far provide the most precise measurements for the decay
$D^0 \rightarrow K^- \pi^+$. They use the same technique by looking at
$D^{*+} \rightarrow \pi^+ D^0$ decays and taking the ratio of the $D^0$ decays
into $K^- \pi^+$ to the number of decays with only the $\pi^+$ from the $D^{*+}$
decay detected. The dominant systematic uncertainty is the background level in
the latter sample. In both experiments, the systematic errors exceed the
statistical errors. 
\begin{table}[ht]
\begin{tabular}{lcc}
\hline
    \multicolumn{1}{l}{\bf }
  & \multicolumn{2}{c}{\bf BR fractional error \%} \\
    \multicolumn{1}{l}{\bf Decay Mode}
  & \multicolumn{1}{l}{\bf PDG 2000}
  & \multicolumn{1}{l}{\bf CLEO-c~\cite{bib:cleoc}}   \\
\hline
$D^0 \rightarrow K \pi$    & 2.4      & 0.6                       \\
$D^+ \rightarrow K \pi \pi$  & 7.2      & 0.7                       \\
$D_s \rightarrow \phi \pi$ & 25       & 1.9                       \\
\hline
\end{tabular}
\caption{Expected branching fractional errors for hadronic decay modes~\cite{bib:cleoc}}
\label{tab:hadronic}
\end{table}
The $D^+$ absolute branching ratios are determined by using fully reconstructed
$D^{*+}$ decays, comparing $\pi^0 D^+$ with $\pi^+ D^0$ and using isospin
symmetry. Hence, this rate cannot be determined any better than the
absolute $D^0$ decay rate using this technique. 
The $D^+_s$ absolute branching ratios are determined by comparing fully reconstructed $B\to D^{(*)}D_s^{*+}$ to the partially reconstructed $B\to D^{(*)}D_s^{*+}$ requiring only the $\gamma$ from the $D_s^{*+}$ decay. Here the dominant systematic uncertainty is due to the background shape in the partially reconstructed sample.
By using $D^0\bar{D}^0$, $D^+D^-$ and $D_s^+D_s^-$ decays, and tagging both $D$ mesons,
the background can be reduced to almost zero and the branching ratio fractional
error can be improved significantly (see Table \ref{tab:hadronic}).

\subsection{Exotic and Gluonic Matter \label{sec:glue}}
The approximately one billion $J/\psi$ produced at CLEO-c will be a
glue factory to search for glueballs and other glue-rich states via
$J/\psi \rightarrow gg \rightarrow \gamma X$ decays. The region of
$1 < M_X < 3$~GeV/$c^2$ will be explored with partial wave analyses for
evidence of scalar or tensor glueballs, glueball-$q\bar{q}$ mixtures, exotic
quantum numbers, quark-glue hybrids and other new forms of matter predicted by
QCD. This includes the establishment of masses, widths, spin-parity quantum
numbers, decay modes and production mechanisms for any identified states, a
detailed exploration of reported glueball candidates such as
the scalar states $f_0(1370)$, $f_0(1500)$ and $f_0(1710)$, and
the examination of the inclusive photon spectrum $J/\psi \rightarrow \gamma$X
with $<$ 20 MeV photon resolution and identification of states with up to 100
MeV width and inclusive branching ratios above $1 \times 10^{-4}$. 


In addition,
spectroscopic searches for new states of the $b\bar{b}$ system and for exotic
hybrid states such as $cg\bar{c}$ will be made using the 4~fb$^{-1}$
$\Upsilon(1S)$, $\Upsilon(2S)$, $\Upsilon(3S)$ and $\Upsilon(5S)$
data sets. Analysis of
$\Upsilon(1S) \rightarrow \gamma X$ will play an important role in verifying
any glueball candidates found in the $J/\psi$ data.

\subsection{Charm Beyond the Standard Model}
CLEO-c will have the opportunity to probe for physics beyond the Standard
Model. Three highlights - rare charm decays, $D^0-\bar{D}^0$-mixing and $CP$ 
violation 
- are discussed in the following sections.

\noindent{\bf Rare Charm Decays:}
Rare decays of charmed mesons and baryons provide ``background-free''
probes of new physics effects. In the framework of the Standard Model (SM)
these processes occur only at one loop level. SM predicts vanishingly small
branching ratios for processes such as $D \to \pi/K^{(*)} \ell^+\ell^-$
due to the almost perfect GIM cancellation
between the contributions of strange and down quarks. 
This causes the SM predictions for these
transitions to be very uncertain.
In addition, in many
cases annihilation topologies also give sizable contribution. 
Several
model-dependent estimates exist indicating that the SM predictions for these
processes are still far below current experimental sensitivities~\cite{bib:rare1,bib:rare2}. 

\noindent{\bf ${\bf D^0-\bar{D}^0}$ Mixing:}
Neutral flavor oscillation in the $D$ meson system is highly suppressed
within the Standard Model. The time evolution of a particle produced as a $D^0$
or ${\overline D}^0$, in the limit of $CP$ conservation, is governed by four parameters:
$x=\Delta m/\Gamma$, $y=\Delta \Gamma/2\Gamma$ characterize the mixing matrix, 
$\delta$ the relative strong phase
between Cabibbo favored (CF) and doubly-Cabibbo suppressed (DCS) amplitudes and 
$R_D$ the DCS decay rate relative to the CF decay rate~\cite{bib:asner}. 
Standard Model based predictions for $x$ and $y$, as well as a variety of non-Standard 
Model expectations, span several orders of magnitude~\cite{bib:Nelson}.
It is reasonable to assume that $x\approx y \approx 10^{-3}$ in the Standard Model.
The mass and width differences $x$ and $y$ can be measured in a variety of ways.
The most precise limits are obtained by exploiting the time-dependence of 
$D$ decays~\cite{bib:asner}. Time-dependent analyses are not feasible at 
CLEO-c; however,
the quantum-coherent $D^0{\overline D}^0$ state provides time-integrated sensitivity 
to $x$, $y$ at ${\cal O}(1\%)$ level and $\cos\delta\sim 0.05$~\cite{bib:cleoc,bib:ddmixing2}. Although CLEO-c does not have sufficient sensitivity to observe Standard
Model charm mixing the projected results compare favorably with current experimental results; 
see Fig.~1 in Ref.~\cite{bib:asner}.

\noindent{\bf ${\bf CP}$ Violation:}
Standard Model $CP$ violation is strongly suppressed in charm. While theoretical
predictions have significant uncertainties,
Standard Model predictions for the rate of $CP$ violation in charm mesons are as large
as 0.1\% for $D^0$ decays and as large as 1\% for certain $D^+$ and $D^+_s$ decays~\cite{bib:Buccella}.

The production process
$e^+ e^- \rightarrow \psi(3770) \rightarrow D^0 \bar{D^0}$
produces an eigenstate of $CP+$, in the first step, since the $\psi$(3770) has
$J^{PC}$ equal to $1^{--}$. Now consider the case where both the $D^0$ and
the $\bar{D^0}$ decay into $CP$ eigenstates. Then the decays
$\psi(3770) \rightarrow f^i_+ f^j_+ ~~or~~ f^i_- f^j_-$
are forbidden, where $f_+$ denotes a $CP+$ eigenstate and $f_-$ denotes a $CP-$
eigenstate. This is because
$CP(f^i_{\pm} ~f^j_{\pm}) = (-1)^\ell = -1$
for the $\ell = 1 ~\psi$(3770).
Hence, if a final state such as ($K^+K^-$)($\pi^+\pi^-$) is observed, one
immediately has evidence of $CP$ violation. Moreover, all $CP+$ and $CP-$ 
eigenstates
can be summed over for this measurement. The expected sensitivity to direct $CP$
violation is $\sim 1\%$.
This measurement can also be performed at higher energies where the final
state $D^{*0} \bar{D^{*0}}$ is produced. When either $D^*$ decays into a
$\pi^0$ and a $D^0$, the situation is the same as above. When the decay is 
$D^{*0} \rightarrow \gamma D^0$ the $CP$ parity is changed by a multiplicative
factor of -1 and all decays $f^i_+ f^j_-$ violate $CP$~\cite{bib:CP}. Additionally, $CP$
asymmetries in $CP$ even initial states depend linearly on $x$ allowing sensitivity to
$CP$ violation in mixing of $\sim 3\%$~\cite{bib:cleoc}.

{\noindent \bf Dalitz Plot Analyses:} A Dalitz plot analysis of multibody final states measures amplitudes and phases
rather than the rates and so may provide greater sensitivity to $CP$ violation.
In the limit of $CP$ conservation, charge conjugate decays will have the same
Dalitz distribution. Although the $D^+$ and $D_s^+$ decays are self-tagging,
there have been no reported Dalitz analyses that search for $CP$ violation with
charged $D$'s. The decay $D^0 \to K_S \pi^+\pi^-$ proceed through intermediate
states that are $CP+$ eigenstates, such as $K_S f_0$, $CP-$ such as $K_S\rho$ and flavor eigenstates such as $K^{*-}\pi^+$~\cite{bib:asner2}. 
It is noteworthy that for uncorrelated $D^0$
the interference between $CP+$ and $CP-$ eigenstates integrates to zero across the 
Dalitz plot but for correlated $D$ the interference between $CP+$ and $CP-$ eigenstates
is locally zero. The Dalitz plots for $\Psi(3770) \to D^0{\overline D}^0 \to f_+K_S\pi^+\pi^-$ and  $\Psi(3770) \to D^0{\overline D}^0 \to f_-K_S\pi^+\pi^-$ will be
distinct and the Dalitz plot for the untagged sample  $\Psi(3770) \to D^0{\overline D}^0 \to X K_S\pi^+\pi^-$ will be distinct from that observed with uncorrelated $D$'s from continuum production at $\sim 10$~GeV~\cite{bib:asner2}.
The sensitivity at CLEO-c to $CP$ violation with Dalitz plot analyses has not yet been evaluated.

\section{Summary}
The high-precision charm and quarkonium data will permit a broad suite of
studies of weak and strong interaction physics as well as probes of
new physics. In the threshold charm sector
measurements are uniquely clean and make possible the unambigous determinations
of physical quantities discussed above. 
The advances in strong interaction calculations enabled by CLEO-c will allow
advances in weak interaction physics in all heavy quark endeavors and in future
explorations for physics beyond the Standard Model.


\bibliographystyle{aipproc}   

\bibliography{Beauty2003}
\hyphenation{Post-Script Sprin-ger}

\end{document}